\documentclass{svmult}

\usepackage{makeidx}
\usepackage{graphicx}
\usepackage{multicol}
\usepackage{placeins}
\makeindex

\begin{document}

\titlerunning{GraphStream: Complex Systems and Dynamic Graphs}
\title*{GraphStream: A Tool for bridging the gap between Complex Systems and Dynamic Graphs}
\author{Antoine Dutot\and
        Fr\'ed\'eric Guinand\and
        Damien Olivier\and
        Yoann Pign\'e \thanks{The authors names are sorted alphabetically.}
}
\institute{LITIS -- FST,
        Universit\'e du Havre,
        25 Rue Philippe Lebon,
        BP 540, 76058 Le Havre cedex,
        France
        \texttt{antoine.dutot@univ-lehavre.fr}
}
\maketitle

\abstract{
The notion of complex systems is common to many domains, from Biology to Economy, Computer Science,
Physics, etc. Often, these systems are made of sets of entities moving in an evolving environment. 
One of their major characteristics is the emergence of some global properties stemmed from local
interactions between the entities themselves and between the entities and the environment. The structure of
these systems as sets of interacting entities leads researchers to model them as graphs. However,
their understanding requires most often to consider the dynamics of their evolution. It is indeed
not relevant to study some properties out of any temporal consideration. Thus, dynamic graphs seem
to be a very suitable model for investigating the emergence and the conservation of some
properties.

\emph{GraphStream} is a Java-based library whose main purpose is to help researchers and developers
in their daily tasks of dynamic problem modeling and of classical graph
management tasks: creation, processing, display, etc. It may also be used, and
is indeed already used, for teaching purpose.  
\emph{GraphStream} relies on an event-based engine allowing several event sources. Events
may be included in the core of the application, read from a file or received from an event
handler.
}

%


%
%
%
\section{From Complex Systems to Dynamic Graphs}

Biology, Physics, Social Sciences, Economy, \textit{etc.} yield all examples
of systems where a large number of entities, homogeneous or heterogeneous,
mutually interact.  
The emergence of global properties leads the community to qualify these
Systems as Complex.
This emergence comes from the dynamic of interactions.
This dynamic is responsible not only for the building of the properties,
which may be consider as a morphogenic process, but also for its maintaining,
comparable to a morphostatic process, and as such it should be present in any
relevant model.   
So, regarding Complex Systems, dynamic can be observed at two levels: at the
level of the system formation and at the level of system evolution. 
Both processes may be distinct or may be the same, as Web graphs for
instance. 
To sum up, we may transpose Dobzhansky's famous maxim: 
{\em Nothing in biology makes sense, except in the light of Evolution} to Complex Systems: 
{\em Nothing in Complex Systems makes sense, except in the light of Dynamic of Interactions}.

Then, performing a relevant study of such Complex Systems requires
a dynamic representation, and dynamic graphs impose themselves as the more
suitable way of modeling them. 

\begin{figure}[h]
        \centering
                \includegraphics[width=0.7\textwidth]{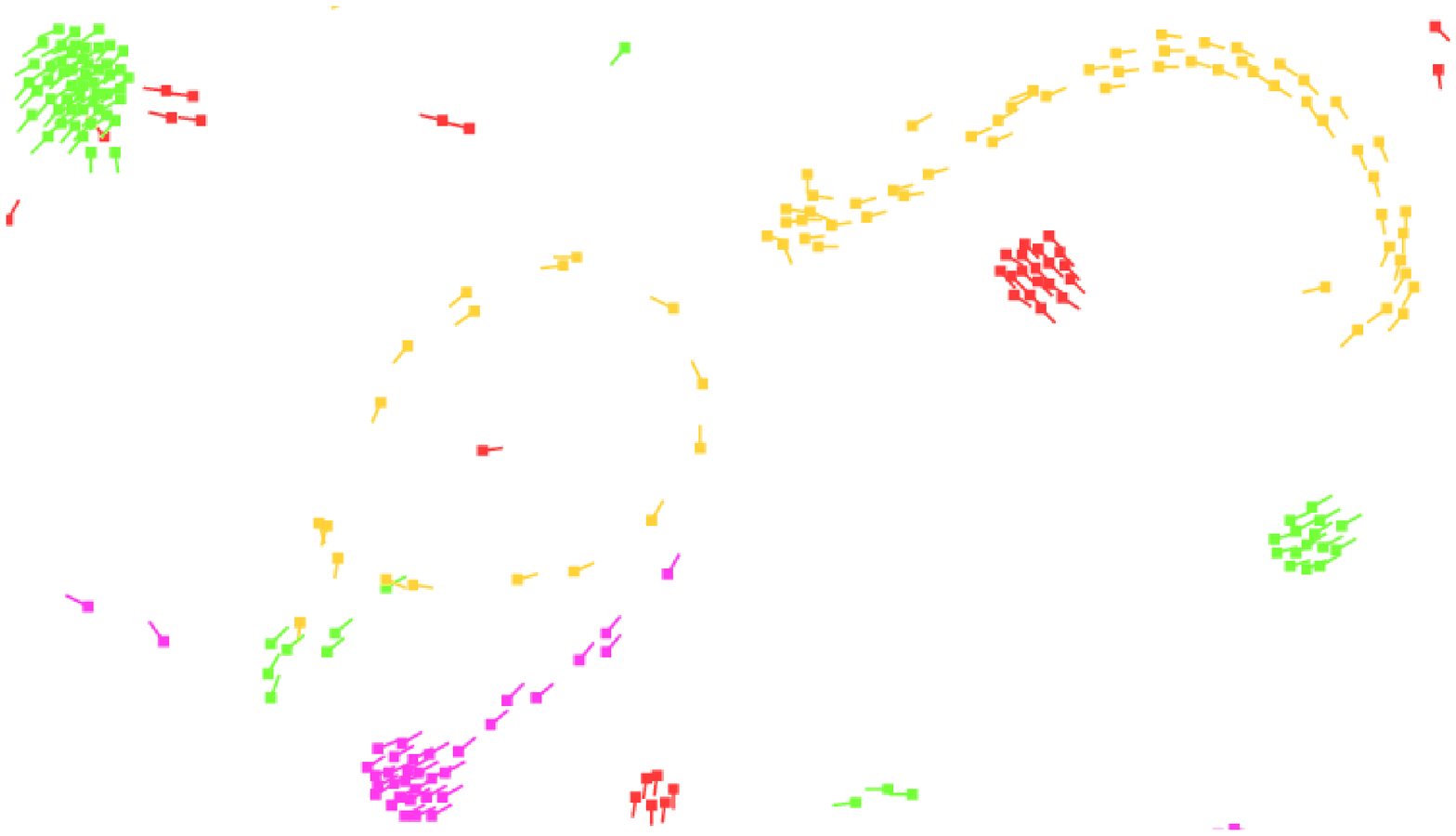}
                $\quad$
                \includegraphics[width=0.6\textwidth]{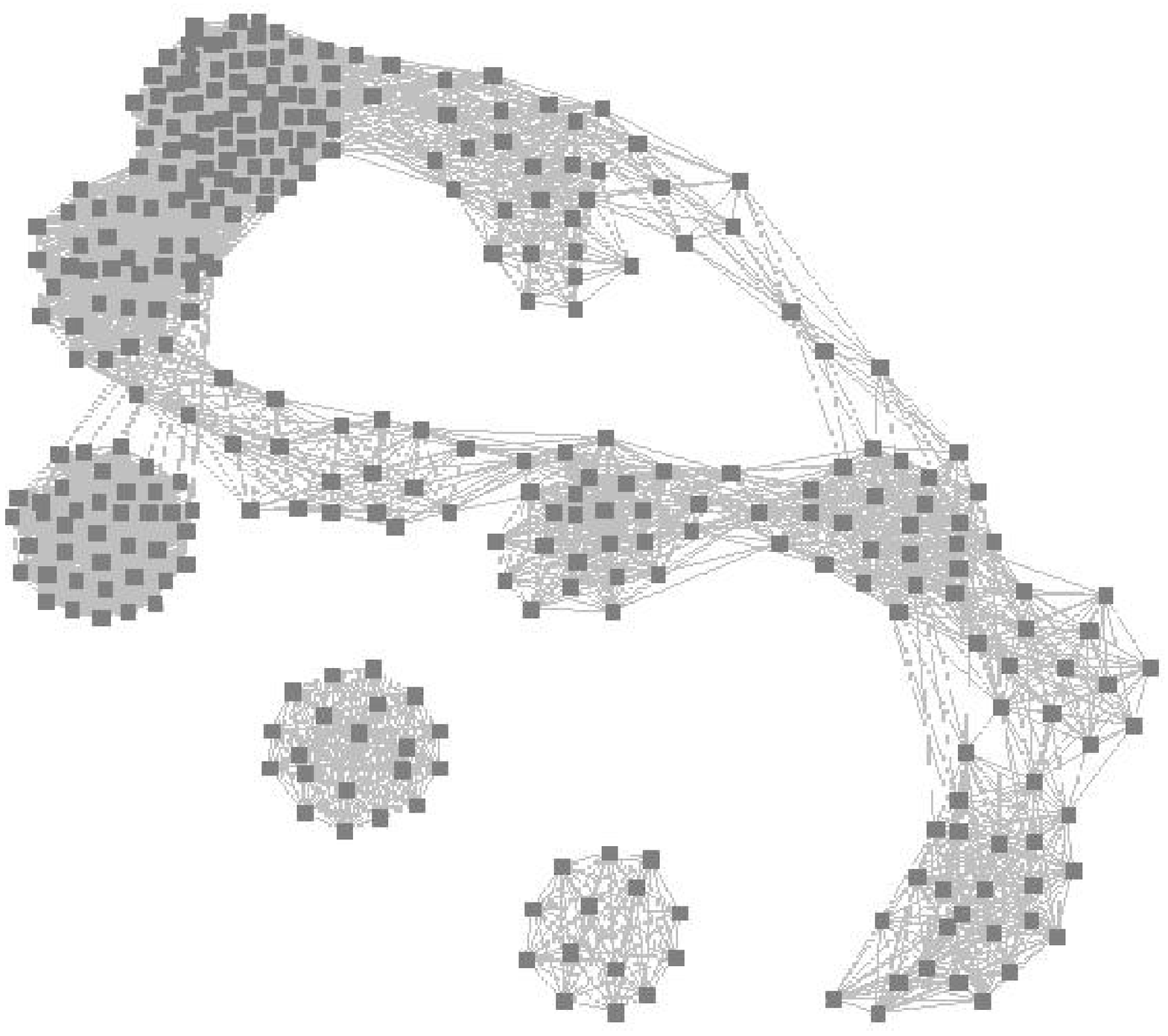}
        \caption{This picture represents one zoom in part of a boids-based simulation (on the
                top) and its representation with a dynamic graph (on the
                bottom). 
                On the left part, a snapshot of the simulation is proposed. 
                The boids (interacting entities) can only perceive and
                interact with fellow creatures located in a limited area.
                During the process, some groups appear corresponding to global
                structures obtained from local interactions.
                On the right part of the figure, the graph represents at a
                given moment the interaction graph corresponding to the
                state of the system made of boids.}
        \label{fig:boids}
\end{figure}

Figure \ref{fig:boids} illustrates how the main processes leading to the
formation and the maintaining of groups within a system composed of many
entities in interaction may be observed using a dynamic graph model.
In this representation, the vertices of the graph represent entities and the
edges represent interactions.  
During the evolution of the system, both kinds of elements, vertices and
edges, may be subject to variations and their characteristics as well. 

\FloatBarrier


 


A Dynamic Graph $G(t) = \left( V(t), E(t) \right)$ where $t$ represents the
time, is composed of:
\begin{itemize}
        \item a set of vertices $V(t)$. This set may change with $t$. Each
          vertex $v$ owns a set of characteristics $C_v(t)$ that may
          vary with time. 
        \item a set of edges $E(t)$. $E(t)$ may change with $t$.
          Each edge $e_{ij} = {v_i,v_j}$ is defined as a pair of vertices, and
          owns a set of characteristics $C_{e_{ij}}(t)$ that may vary with
          time. 
\end{itemize}

Therefore a dynamic graph $G(t)$ may change with the time $t$. According to
the underlying modeled system, a dynamic graph may be considered as a family
of graphs or as a complete graph where, according to $t$ some vertices or
edges can be zero weighted. 
There exist however another way of defining a dynamic graph, based on 
a discrete-time representation, and thus very well suited for most
simulations.

\begin{figure}[h]
        \centering
                \includegraphics[width=\linewidth]{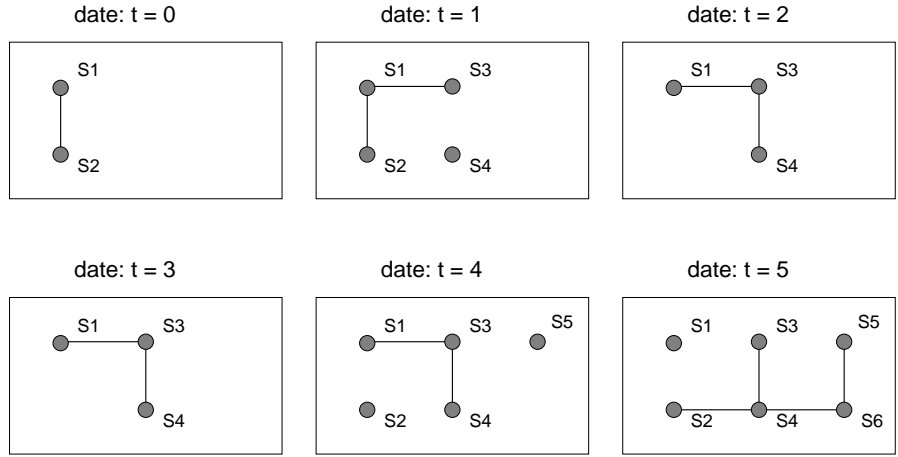}
        \caption{Snapshots of a Dynamic Graph on the time interval
                $[0,5]$. During a period of time the graph may not change as
                it is the case during interval $[2,3]$. Some vertices may
                appear and disappear and appear again, as illustrated by
                vertex $s_2$.}
        \label{fig:dt-model}
\end{figure}

A dynamic graph can be defined as a finite or infinite ordered set of couples
$(\mbox{date}, \{\mbox{events}\})$. 
Every set of events may modify the graph structure/composition/topology and/or
characteristics of some graph elements.
The different snapshots of Figure \ref{fig:dt-model} correspond to the series
of sets of events listed below:

\begin{itemize}
\item[] (0,\{$v_1$ creation, $v_2$ creation, $(v_1,v_2)$ creation\})
\item[] (1,\{$v_3$ creation, $v_4$ creation, $(v_1,v_3)$ creation\})
\item[] (2,\{$v_2$ suppression, $(v_3,v_4)$ creation\})
\item[] (4,\{$v_2$ creation, $v_5$ creation\})
\item[] (5,\{$v_6$ creation, $(v_5,v_6)$ creation, $(v_4,v_6)$ creation, $(v_2,v_4)$ creation, $(v_1,v_3)$ suppression\})
\end{itemize}

This approach was chosen for the implementation of \emph{GraphStream} as
explained in the forthcoming Section. 


 


\section{Graphs streams}

\emph{GraphStream} \cite{GraphStream} is a Java library whose purpose is to
create and manipulate dynamic graphs in order to study them and to use them in
simulations.  
The whole library has been devised with the idea that graph will evolve during
time. 
It is also naturally well suited for the manipulation of "static" graphs.

The central notion in \emph{GraphStream} is that graphs are not static
objects, but streams of events, and these streams can flow from producers
(simulations, graph generators, graph readers, the web, etc.) passing by
filters (a graph structure, a converter, etc.) to outputs (graph writers, a 
network, a display, etc). There also exist a file format called {\sc dgs}
allowing the expression of graph changes as series of events that can be read
by \emph{GraphStream}.      

Around this notion, \emph{GraphStream} provides several tools.

\subsection{Packages}

\begin{description}
         \item{\textbf{Graph Representation Tools}} The main package provided by \emph{GraphStream}
                allows the representation of graphs in memory. It is made of a graph interface and several
                implementation fulfilling different needs or usages. All share the property to handle
                nicely the dynamics of what they represent. Some of them are able to store arbitrary data
                in addition to the sole graph dynamic topology, whereas others are made to be small and
                fast. The library does not force the creation of a graph structure in memory. Several of
                its tools can handle a flow of events without ever constructing a whole representation of
                it in memory. For example it is possible to write a filter that will transform a very large
                graph written on disk into another graph on disk without necessitating a growing amount of
                memory.
         \item{\textbf{Graph Generation Tools}} Most often, a simulation or observation tool will be
                used to produce graph events. However the library proposes a bunch of graph generators going
                from regular graphs like grids and tores, to scale-free graphs, random graphs, etc. It makes
                it easy to test algorithms on graphs having specific properties.
         \item{\textbf{Graph I/O Tools}} The library can produce streams of events from several file
                formats read either from a file system or from the internet. This flow of event is not a
                graph and therefore, it offers the ability to process graphs that are larger than the
                available memory if no global representation is needed. Similarly, it is possible to feed a
                stream of graph events toward a file to save it in several formats. A specific format as
                been create to manage the complete set of graph events, not only the fact that nodes and
                edges exist, but also that they evolve.
         \item{\textbf{Graph Theory Tools}} The library offers several well known algorithms on graphs
                and tries, where possible, to propose version of them that can handle the graph dynamics
                (this is the part of the library that still needs the more work, though, and can
                provide some interesting problems). For example, the connected components count algorithm not
                only is able to count the number of connected components of the graph, but is able to track
                it without having to recompute all the connected components count as the graph evolves. An
                important field of research named re-optimization aims at maintaining a previous result and 
                recomputing parts of a solution when a changes occur in the environment. The idea being that
                the re-optimization of a previously obtained result is more efficient than re recomputing of
                the solution from scratch. An example of successful re-optimization algorithm can be found
                with the Dynamic Shortest Path Problem \cite{Demetrescu2001}. 
         \item{\textbf{Graph Visualization Tools}} The library is equipped with display capabilities. It
                makes it easy to display any graph structure either on the local machine or even on a
                distant one. The viewer understands various style attributes (colors, dashes, arrows, labels
                on nodes and edges, icons). It is equipped with an optional automatic layout engine that
                can compute optimal node positions in order to make the graph as visible as possible (by
                avoiding edge crossing). This engine handles the graph dynamics and constantly reorganises
                as the graph evolves.
\end{description}

\subsection{Example}

\begin{figure}[h]

\begin{verbatim}
import org.miv.graphstream.graph.*;
 
public class Easy1
{
    public static void main( String args[] )
    {  
        Graph graph = new Graph( false, true ); 
        graph.addEdge( "AB", "A", "B" ); 
        graph.addEdge( "BC", "B", "C" ); 
        graph.addEdge( "CA", "C", "A" ); 
        graph.display();
    }  
}
\end{verbatim}

        \centering
                \includegraphics[width=0.35\textwidth]{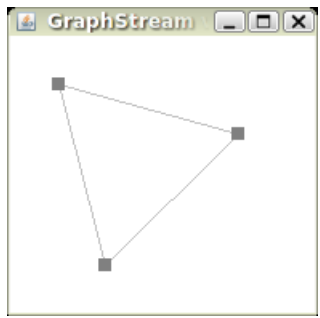}
        \caption{The GraphStream viewer.}
        \label{fig:viewer}
\end{figure}

Creating and visualizing a graph using \emph{GraphStream} is more than easy. 
Indeed, the following piece of code creates a graph made of 3 vertices and 3
edges and open a window for visualizing it, the whole in only 8 lines of code.
The result is shown Fig.~\ref{fig:viewer}). 
The positions of vertices are automatically computed in order to enhance the
display. This helper can however be avoided by replacing the "true" boolean
value by "false" in the parameters of Graph() constructor. 

\FloatBarrier

\subsection{Suitability for Complex Systems Simulations}


\emph{GraphStream} is built upon an event-based engine and its whole
conception is object-oriented, each element of the graph (nodes and edges) but
also each attribute that qualify these elements are objects. 
These features allow an easy prototyping and simulation of systems made
of sets of interacting entities.
In particular:

\begin{itemize}
\item discrete-time simulations are easy to implement by associating
  an event to each time increment,
\item the object approach allows a decentralized point of view that prevails
  over centralized approaches for Complex Systems modeling since interactions
  between entities are mainly characterized by their locality,
\item finally, the library offers a lot of classical graph theoretic
  algorithms that may help for the analysis of interactions graphs. 
\end{itemize}

In order to illustrate these claims, in the sequel an example of application
in the fields of wireless telecommunication networks is developed.

\section{Case study: Wireless Mobile Networks}

\subsection{Mobile Ad Hoc Networks}

Nowadays almost all notebooks are equipped with wireless communication materials.
These devices, that are also present in many smartphones and PDAs,
are able to perform communications in two modes: {\em infrastructure} or {\em
  ad hoc}.   
In the former mode, the stations (computers, mobiles phones, PDAs, etc)
communicate with each others via an access point. This access point manages the
communications in the Wireless LAN and acts as a gateway for communications
outside of the local network. This mode is usually used at home or in public
places.  
In ad hoc mode, no access point is necessary for performing communications,
stations "simply" broadcast their messages in their neighborhood. 
Such networks are called MANETs (Mobile Ad hoc NETworks) and when the
assumptions allow the network to be partitioned, we call them DT-MANETs,
meaning {\em Delay Tolerant} or {\em Disruption Tolerant} MANETs.  
Although it is not commonly used, ad hoc mode presents some advantages. In
particular neither infrastructure nor network administration are required, and
this mode is especially well-suited for mobility \cite{Hogie2006c}.
However, classical network problems like broadcasting or routing have to be
revisited because stations may move in their environment, they may also be
turned on or off at any moment such that the topology of the network changes
while the operation is performed. 
The complexity of these systems comes from the mobility of stations. 
Indeed, as stations move according to the behavior of humans, situations
occurring in real life (traffic jams, groups of people at one moment at one
place, etc) also occur within such networks.
The suitability of \emph{GraphStream} for that task is obvious since the
communication network is dynamic and can be modeled with a 
\emph{dynamic graph} and from a station's point of view the network as a whole
does not exist, each station only knows the stations that are within their
neighborhood, that is the network as a whole is only a representation, it is
actually decentralized. 

\subsection{Implementation choices}

In the context of this example we are interested in testing algorithms that
build spanning forests within DT-MANETs. 
Indeed, spanning forests may be used as a basic structure for different
operations like routing or broadcasting messages between stations.
There exist at least two ways for performing such a task.
The first option consists in relying on an existing DT-MANET simulator and to
gather for each time slot events like: station $s_i$ turns on, stations $s_j$
and $s_k$ are in mutual communication range, etc. 
The second option consists in rebuilding a simulator from scratch using 
\emph{GraphStream}.
Because we already have at our disposal such a simulator ({\sc Madhoc}
\cite{Hogie2007}), we choose the first solution.

The whole process for testing algorithms first recover events from a run of 
{\sc Madhoc} and keep them in a file. 
The IO part of \emph{GraphStream} is used to read this file and generate the
events accordingly by associating to each station a node and to each
communication link an edge. 
\emph{GraphStream} is also used to simulate the Spanning Forest
Algorithm. So as to benefit the Object oriented approach of the library, the 
algorithm was modeled consequently. 
Before entering in more details, let us described the method for building and
maintaining a spanning forest in a dynamic graph.

\subsection{Building and Maintaining a Spanning Forest}

The Spanning Forest Algorithm \cite{casteigts_06} is a mechanism based on
a set of rules that are applied on each station of the network in a fully
decentralized way. 
The application of these rules, locally by each station, results in the
marking of a set of edges of the graph that form a spanning forest. 
In the best case, the process leads to one single tree for each connected
component of the graph. 
Since the process is repeated (the rules) all the time, it adapts to the
dynamics of the network. 
The root of each tree corresponds to a node owning a \emph{token}. 
Each token moves along the constructed tree.   
After a local synchronization process with its neighbors, each station
executes 4 rules:    

\begin{itemize}
\item Rule 1: An edge belonging to the tree has been removed. At this moment,
  the tree is split in two. This rule applies to the node belongs to the part
  of the tree where there is no token (no root). So the node creates a token
  on itself (a new root).  
\item Rule 2: An edge is removed on the tree and the node is part of the tree
  where the token is.  
\item Rule 3: Two nodes with tokens meet. One of the two tokens disappears and
  the two trees merge.   
\item Rule 4: This last rule makes the token move along the tree. 
\end{itemize}

\begin{figure}[h]
        \centering
                \includegraphics[width=\textwidth]{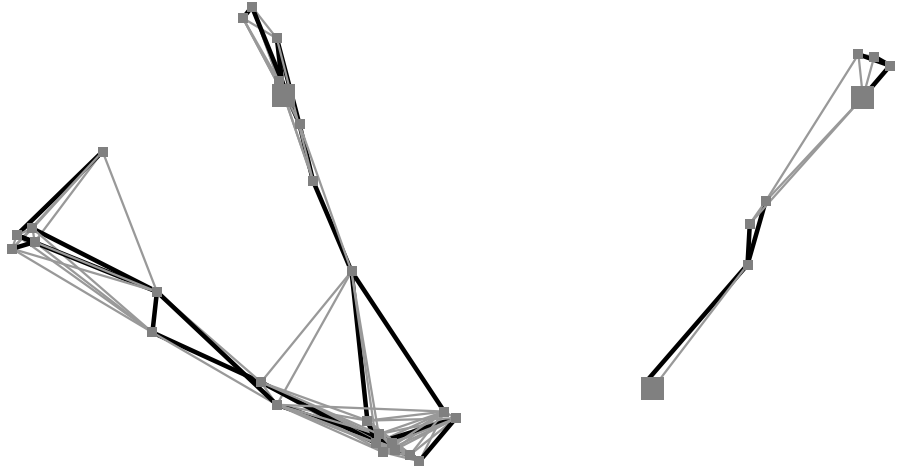}
        \caption{A graphical view of a MANET. Black thick links are colored by
                the Spanning Forest Algorithm and belong to the forest. Bigger
                node are the ones owning a \emph{token}, they are the roots or
                the spanning trees.} 
        \label{fig:SpanningForestScreenshot}
\end{figure}

For the purpose of the simulation with \emph{GraphStream}, an agent-based
model as been used. The 2 first rules of the above model can be implemented as
a reaction to edge events. 
Rule 4 is pro-active and defines the movement of the token. 
Rule 3 is executed as a consequence to rule 4. 
The concept of token is modeled as an agent that traverses the
network. 
Fig. \ref{fig:SpanningForestScreenshot} shows a graphical representation of
such a network with the Spanning Forest Algorithm running on it.
  
\subsection{Measurements}

The real gain in using \emph{GraphStream} with such an algorithm is that it
allows both the decentralized point of view necessary for the algorithm and
the centralized measurements that give an interesting analysis of its behavior.  

\begin{figure}[!h]
        \centering
                \includegraphics[width=\textwidth]{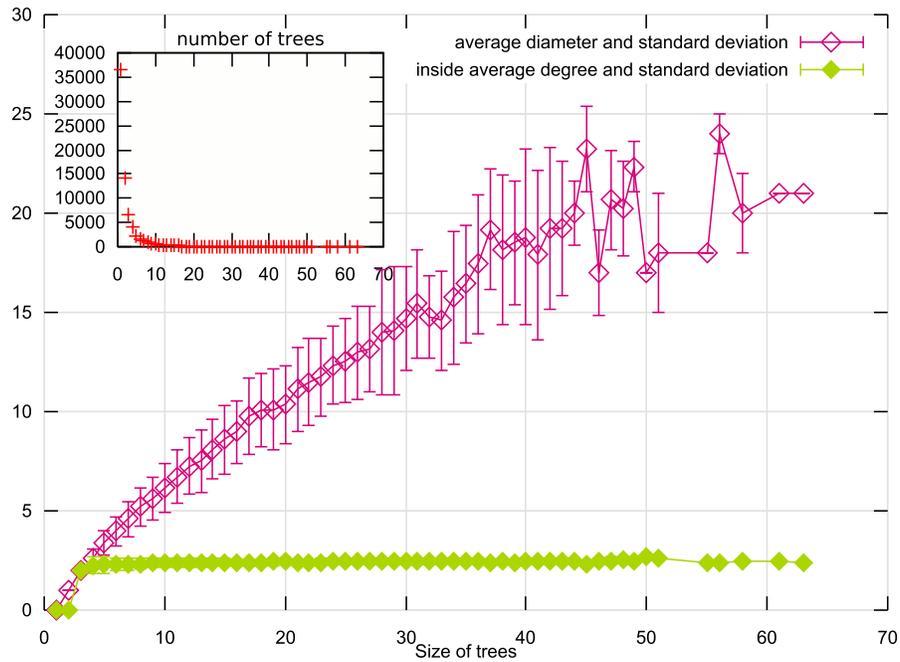}
        \caption{ Evolution of the average diameter and the average degree of
                inner nodes of trees. The small top left chart
                represents the number of trees w.r.t. their size.} 
        \label{fig:measurements}
\end{figure}

An example of the measurements made thanks to the library can be observed on
Figure \ref{fig:measurements}. 
This chart represents the evolution of some measures made on the trees
constructed with the defined algorithm. 
This evolution is observed here according to the size (number of nodes) of the
trees.   

\FloatBarrier
\section{Conclusion}

A new dynamic graph representation and manipulation library was presented in
this paper. The originality of the approach rests in the way dynamics is
handled in graphs.  
The idea of a stream of events for the dynamics and the general object
oriented outlook of the library give it a special ability to be used as a
distributed and dynamic environment simulator. \emph{GraphStream} also
advertises more classical tools for analyzing static and dynamics graphs.      

The usefulness of the library was shown with the modeling and the simulation
of a decentralized algorithm. In this simulation \emph{GraphStream} was used
as a simulator. It was also use to perform an analyze of the behavior of the
algorithm with graph theory measurements.


\bibliographystyle{abbrv}
\bibliography{GraphStream}

\end{document}